\documentclass[aip,amsmath,amssymb,reprint]{revtex4-1}
\usepackage{dcolumn}
\usepackage{bm}
\usepackage[utf8]{inputenc}
\usepackage[T1]{fontenc}
\usepackage{mathptmx}
\usepackage[pdftex]{graphicx}
\usepackage{latexsym,amsmath,verbatim}
\usepackage{xcolor}
\usepackage{hyperref}
\usepackage{lipsum}
\usepackage[english]{babel}

\newcommand{\av}[1]{\left\langle {#1} \right\rangle}

\newcommand{\be}{\begin{equation}}
\newcommand{\ee}{\end{equation}}
\newcommand{\ep}{\epsilon}

\graphicspath{{../}}

\begin{document}

%\preprint{AIP/123-QED}

\title[Message-passing theory for cooperative epidemics]{Message-passing theory for cooperative epidemics}

\author{Byungjoon Min} 
 \email{min.byungjoon@gmail.com}
 \affiliation{Department of Physics, Chungbuk National University, Cheongju, Chungbuk 28644, Republic of Korea}
\author{Claudio Castellano}
 \email{claudio.castellano@roma1.infn.it} 
 \affiliation{Istituto dei Sistemi Complessi (ISC-CNR), Via dei Taurini 19, 
I-00185 Roma, Italy}
\date{\today}

%\pacs{89.65.-s}{Social and economic systems}
%\pacs{89.75.Hc}{Networks and genealogical trees}
%\pacs{89.20.Hh}{World Wide Web, Internet}

\begin{abstract}
The interaction among spreading processes on a complex network 
is a nontrivial phenomenon of great importance. 
It has recently been realized that cooperative effects among 
infective diseases can give rise to qualitative changes in the
phenomenology of epidemic spreading, leading for instance to abrupt
transitions and hysteresis.
Here we consider a simple model for two interacting pathogens on
a network and we study it by using the message-passing approach.
In this way we are able to provide detailed predictions for the
behavior of the model in the whole phase-diagram for any given 
network structure. Numerical simulations on synthetic networks (both homogeneous 
and heterogeneous) confirm the great accuracy of the theoretical results.
We finally consider the issue of identifying the nodes where
it is better to seed the infection in order to maximize the probability
of observing an extensive outbreak. The message-passing approach provides 
an accurate solution also for this problem.
\end{abstract}

\maketitle

\begin{quotation}
How to predict and control mutually cooperative epidemics on a networked system
is an outstanding problem of much interest. Previous attempts in this problem 
have assumed well-mixed populations and focused on the mean-field analysis.
A full analysis of the cooperativity in epidemics at the level of single node
is still lacking. In this work, we provide a
message-passing approach in order to fully analyze cooperative epidemics 
on complex networks. We confirm the great accuracy of our theory with numerical 
simulations for both homogeneous and heterogeneous network structures.
In addition, we reveal how to identify precisely influential spreaders in 
cooperative epidemics by using the message-passing equations.
\end{quotation}

\section{Introduction}

Many types of spreading phenomena, such as the propagation of
infectious diseases or the diffusion of information or fads, 
are often strongly influenced by cooperativity effects.
For instance, the epidemic spreading of a contagious disease can be 
dramatically boosted by the interaction with other infective pathogens. 
This cooperative effect is at the origin of devastating epidemics such as the 
co-occurrence of Spanish flu and pneumonia in 1918~\cite{Morens2008,Brundage2008}
and the concurrent outbreaks of HIV/AIDS and a host of other 
diseases such as tuberculosis~\cite{Sulkowski2008,Pawlowski2012}.
While the theoretical study of the epidemic spreading of a single-pathogen 
on complex topologies has a long and successful history~\cite{PastorSatorras2015}, 
intertwined with the theory of percolation and dynamical processes on 
networks~\cite{Newmanbook,Dorogovtsev2008}, 
the spreading of cooperative epidemics started to attract attention only recently.

Previous research has shown that cooperativity can give rise to violent 
outbreaks, which are the signature of an abrupt transition between the 
disease-free and the epidemic phase~\cite{Chen2013,Chen2015,Chen2016,Janssen2016,Kahngpre2017}.
In addition to the discontinuous change, cooperative epidemic models
are often characterized by the so-called ``hybrid phase transition'', 
a non-trivial critical phenomenon exhibiting features of both first-
and second-order transitions~\cite{Goltsev2006,MinMaxi2018,Kahng2017,Baek2019}. 
The initial investigations on cooperative epidemics have been extended
to analyze the role of degree heterogeneity~\cite{Cui2017},
clustering~\cite{HebertDufresne2015,Cui2019}, multiplex 
networks~\cite{AzimiTafreshi2016,Wei2016,Liu2018}, 
temporal networks~\cite{Rodriguez2017}, cooperative recurrent 
epidemics~\cite{Chen2017}, 
more than two cooperating pathogens~\cite{Zarei2019}.

In this work, we study the spreading of cooperative epidemics
by extending to the case of two pathogens the message-passing approach 
fruitfully used for single-pathogen dynamics~\cite{Karrer2010,Lokhov2014,Min2018}.
In order to account for the cooperativity between two different 
pathogens, we assume the probability of infection to depend on 
the state of the nodes. Specifically, when a node has already 
been infected with one of the pathogens, the probability that 
the second disease is transmitted to its neighbors through a 
contact grows. Previous investigations of this problem have been based 
on homogeneous and heterogeneous mean-field analyses. 
By suitably defining the transmission probabilities of the
first and of the second disease, it is possible to treat the 
incoming transmission probability for each node as uncorrelated
and in this way to extend the conventional message-passing 
approach to cooperative epidemics.
Within the approach, we obtain a set of 
coupled equations whose numerical solution allows us to fully predict 
and control the effect of cooperativity 
at the level of single node, beyond well-mixed populations.
Next, by using the framework we derive a number of 
analytical predictions for random network ensembles.
Finally, as an important application, 
we tackle the problem of identifying influential spreaders, i.e.,
where an epidemic should be seeded to maximize the probability
and the size of a global epidemic 
outbreak~\cite{Kitsak2010,Lu2016,Radicchi2016,Min2018,Cui2019}. 
The message-passing approach allows us to precisely predict these
quantities for any pair of singly-infected seeds.
Numerical simulations are performed to check all our analytical
predictions on synthetic networks, finding overall an excellent 
agreement.

\section{Model for cooperative epidemics}

The basic model for coinfections has been introduced by
Chen, Ghanbarnejad, Cai and Grassberger (CGCG)~\cite{Chen2013}.
It considers two different pathogens, A and B, each evolving
according to a SIR dynamics. Each node can be in one of three states
with respect to each pathogen. For example, with respect to
pathogen A a node can be susceptible ($S$), infected and
able to spread the infection further ($A$) or recovered ($a$).
The state of a node is specified by indicating in which
of the three possible states the node is with respect to each of 
the two pathogens. There are overall 9 possible individual
states: $S, A, B, AB, Ab, aB, a, b, ab$. Notice that
states denoted by a single letter assume the omitted
letter to be $S$.

In this paper, we propose a slightly different cooperative epidemic
dynamics from the CGCG model. We still consider two different
pathogens and in total 9 possible individual states. The cooperativity
between two pathogens is defined as follows: for an agent infected
with both pathogens, the probability to transmit the infection through
a contact increases. The biological rationale of this modeling is that
an infection with a pathogen may cause an individual to pass from the
latent state to a fully infective one with respect to another
pathogen.  For instance, the infection of HIV is the most important
risk factor for progressing from latent tuberculosis, which is hardly
infectious, to active tuberculosis~\cite{Pawlowski2012}.  In this way
the outgoing infection probability of the second disease
increases when a node has already been infected with the
other disease.  On the other hand, in the CGCG model an 
individual infected with one disease has an increased probability 
to get infected with the second disease.

The crucial advantage of our modified cooperative model from a
theoretical perspective is that the incoming transmission
probabilities become effectively uncorrelated. The incoming
probability does not depend on the current state of the node that can
be infected, rather from the state of the node that attempts to infect.  
This lack of correlation allows us to use a message-passing
approach to predict analytically the steady state behavior of
cooperative epidemics.  In this way it is possible to analyze 
and predict the cooperative
epidemics on structured networks beyond the mean-field 
approximation~\cite{Zarei2019}. 
As it will be shown below, this dynamics exhibits a
behavior perfectly analogous to the original CGCG model. 
The original CGCG dynamics is actually not suitable for a message-passing
approach because incoming transmission probabilities are correlated.
In that dynamics the incoming transmission probability is not only
determined by the state of the source node but strongly depends on the
state of the target node. 
Hence, the incoming transmission probabilities for each node are 
correlated and the message-passing approach cannot be applied to 
the CGCG dynamics.

In practice, the dynamics proceeds as follows. At the beginning
of each discrete time step, a list of all infected nodes is
recorded. Then we select at random from the list an infected node $i$
and one of its neighbors, $j$.
If $i$ is infected with a single-pathogen A (or B) and it is susceptible
with respect to the other, it attempts to infect $j$ with the 
pathogen and the infection is actually transmitted
with probability $p_A$ (or $p_B$).
If $i$ is not susceptible with respect to the other pathogen 
(i.e. it is currently infected or it has been infected in the past
and is now recovered) then the probability of transmission is $q_A$ (or $q_B$).
For a node in state $AB$, it is also possible that during
the same time step $i$ infects $j$ first with pathogen A 
(with probability $p_A$) and then with pathogen B (probability $q_B$)
or vice versa.
After this infection attempt has been repeated for all the nodes
in the list of infected nodes, all of them recover and
the time step is over.
In the rest of the paper we will consider, as in almost all
previous studies, only the symmetric case,
where $p_A=p_B=p$ and $q_A=q_B=q$. 
The investigation of the nonsymmetric case is an interesting issue 
that remains open for future work.
In summary, the state of node $j$ is updated according 
to the following transition rules:
\begin{gather}
A\ +\  S\  \xrightarrow{\text{\quad $p$ \quad}}\							A\ +\ A  \nonumber \\
A\ +\  B\  \xrightarrow{\text{\quad $p$ \quad}}\							A\ +\ AB  \nonumber \\
Ab\ +\  S\ \xrightarrow{\text{\quad $q$ \quad}}\							Ab\ +\ A  \nonumber \\
Ab\ +\  B\ \xrightarrow{\text{\quad $q$ \quad}}\							Ab\ +\ AB  \nonumber \\
B\ +\  S\  \xrightarrow{\text{\quad $p$ \quad}}\							B\ +\ B  \nonumber \\
B\ +\  A\  \xrightarrow{\text{\quad $p$ \quad}}\							B\ +\ AB  \nonumber \\
aB\ +\  S\ \xrightarrow{\text{\quad $q$ \quad}}\							aB\ +\ B  \nonumber \\
aB\ +\  A\ \xrightarrow{\text{\quad $q$ \quad}}\							aB\ +\ AB  \nonumber \\
AB\ +\  S\ \xrightarrow{\text{\quad $pq$ \quad}}\				AB\ +\ AB  \nonumber \\
AB\ +\  S\ \xrightarrow{\text{\quad $\frac{1}{2} p(1-q)$ \quad}}\                                  AB\ +\ A  \nonumber \\
AB\ +\  S\ \xrightarrow{\text{\quad $\frac{1}{2} p(1-q)$ \quad}}\                                  AB\ +\ B  \nonumber \\
A\ \xrightarrow{\text{\quad 1 \quad}}\ a\ \nonumber \\
B\ \xrightarrow{\text{\quad 1 \quad}}\ b\ \nonumber \\
Ab\ \xrightarrow{\text{\quad 1 \quad}}\ ab\ \nonumber \\
aB\ \xrightarrow{\text{\quad 1 \quad}}\ aa\ \nonumber \\
AB\ \xrightarrow{\text{\quad 1 \quad}}\ ab\ \nonumber 
\end{gather}

\section{Message-passing theory}

\subsection{Message-passing equations}

In the symmetric case, by construction the probabilities of the
process do not depend on which of the two pathogens is considered.
We can then define a single function $u_{ij}$, as the probability that,
at the end of the dynamics, node $i$ has not been infected by node $j$ 
with a given pathogen.
In analogy with the case of SIR single-pathogen 
dynamics~\cite{Radicchi2016,Min2018}, 
the message-passing equations for these quantities are
\begin{widetext}
\begin{align}
1-u_{ij} &=  \frac{1}{2} (p+q) \left( 1- \prod_{k \in \partial j \setminus i} u_{jk} \right) \left( 1- \prod_{k \in \partial j} u_{jk} \right) 
+p \left( 1- \prod_{k \in \partial j \setminus i} u_{jk} \right) \left( \prod_{k \in \partial j} u_{jk} \right). 
\label{u} 
\end{align}
\end{widetext}
where $k \in \partial j$ represents the set of neighbors of node $j$
and $k \in \partial j \setminus i$ is the set of neighbors of node $j$
excluding node $i$.  The second term on the r.h.s. takes into account
the case node $i$ is infected with a pathogen by a node $j$ that is
never infected with the other pathogen. The first term accounts
instead for the case $i$ is infected with a pathogen by a node $j$
that at the end is also infected with the other. The factor $1/2$ in $(p+q)/2$
is the probability that the node is infected with the given pathogen
earlier or later than with the other. 
Here we assume that the networks have a locally tree-like 
structure, i.e., they contain no short loops. 
For highly clustered networks, Eq.~\ref{u} cannot guarantee to 
produce accurate probabilities and hence more advanced treatments
\cite{Newman2009,Radicchiprer2016,Cantwell2019} are needed.
In the non-cooperating case, $q=p$, Eq.~(\ref{u}) correctly coincides 
with the message-passing equation for single-pathogen SIR dynamics. 
Note that $u_{ij}=1$ is the trivial fixed point corresponding to 
no outbreak propagation.

\subsection{Derivation of the epidemic threshold} 

In order to find the epidemic threshold, 
let us set $u_{ij}=1-\ep_{ij}$ and expand Eq.~(\ref{u}) 
for small $\ep_{ij}$:
\begin{widetext}
\begin{align}
\ep_{ij} &= \frac{1}{2}(p+q) \left[ 1- \prod_{k \in \partial j \setminus i} (1-\ep_{jk}) \right] \left[ 1- \prod_{k \in \partial j} (1-\ep_{jk}) \right]
+p \left[ 1- \prod_{k \in \partial j \setminus i} (1-\ep_{jk}) \right] \left[ \prod_{k \in \partial j} (1-\ep_{jk}) \right] \nonumber \\ 
& \approx \frac{1}{2}(p+q)  \sum_{k \in \partial j \setminus i} \ep_{jk}  \sum_{k \in \partial j} \ep_{jk} 
+p \left( \sum_{k \in \partial j \setminus i} \ep_{jk} - \frac{1}{2}\sum_{\substack{k,l \in \partial j \setminus i \\ l\ne k}} \ep_{jk} \ep_{jl} \right) \left( 1-\sum_{k \in \partial j} \ep_{jk} \right). 
\end{align}
\end{widetext}

Neglecting second and higher order terms, we obtain 
\begin{align}
\ep_{ij} &\approx p \sum_{k \in \partial j \setminus i} \ep_{jk}.
\end{align}
Defining the $2E \times 2E$ non-backtracking matrix 
$B$~\cite{Hashimoto1989,Karrer2014} 
(where $E$ is the number of edges in the network) 
with elements
\be
B_{i \to j,l \to k}=\delta_{jl}(1-\delta_{ik})
\ee
we can write
\begin{align}
\ep=p B \ep.
\end{align}
An immediate consequence is that 
the epidemic threshold is given by the inverse of the principal
eigenvalue $\Lambda_M(B)$ of the non-backtracking matrix $B$
\be
p_c = \frac{1}{\Lambda_M(B)}.
\label{threshold}
\ee
The parameter $q$ does not appear in Eq.~(\ref{threshold}),
hence $p_c$ coincides with the epidemic threshold for single-pathogen
dynamics.

Keeping terms up to second order, we obtain
\begin{align}
\ep_{ij}& \approx p \sum_{k \in \partial j \setminus i} \ep_{jk} 
+ \frac{1}{2} (q-p) \sum_{k \in \partial j \setminus i} \ep_{jk} \sum_{k \in \partial j} \ep_{jk} -\frac{p}{2}\sum_{\substack{k,l \in \partial j \setminus i \\ l\ne k}} \ep_{jk} \ep_{jl}. 
\label{eq:ep_second}
\end{align}
Depending on $q$, the second order terms in Eq.~(\ref{eq:ep_second}) 
can be either positive or negative.
The transition of the outbreak size at $p_c$ becomes discontinuous
when $q$ becomes larger than a certain threshold $q_c$,
while the transition remains continuous when $q<q_c$.
Note that the ordinary SIR model always produces
a continuous transition since the second order term for the ordinary 
SIR model ($q=p$) is always negative.

\subsection{Outbreak size}

Once the coupled equations~(\ref{u}) 
are solved -- for example, by iteration --
the probabilities that node $i$ has been 
infected with both pathogens, or with one pathogen but not with the other 
are, respectively:
\begin{align}
\rho_{i}^{ab} &=  \left( 1- \prod_{j \in \partial i} u_{ij} \right)^2, \\  
\rho_i^a &=  \left( 1- \prod_{j \in \partial i} u_{ij} \right) \left(\prod_{j \in \partial i} u_{ij}\right).
\end{align}
The average size of cooperative epidemic outbreaks is therefore
\begin{align}
\rho_{ab} 
&= \frac{1}{N} \sum_i 
\left( 1- \prod_{j \in \partial i} u_{ij} \right)^2.
\label{rhoAB}
\end{align}
By the same token, the fraction of nodes infected only by a given 
pathogen is
\begin{align}
\rho_a 
&= \frac{1}{N} 
\sum_i \left( 1- \prod_{j \in \partial i} u_{ij} \right) \left( \prod_{j \in \partial i} u_{ij} \right).
\label{rhoA}
\end{align}

\subsection{Probability of coinfection outbreaks}

Let us determine the probability of coinfection outbreaks 
initiated by two singly-infected seed nodes.
For instance, pathogen A is initially seeded in node $i$
while pathogen B is seeded in node $j$.
A necessary condition for the formation of a global coinfection 
is that each pathogen is individually able to originate an 
extensive single-pathogen outbreak, so that the two outbreaks 
can both infect some nodes. Only after independently evolving
single-pathogen outbreaks have started to overlap,
cooperativity may begin to play its role.
To assess the probability of a single-pathogen outbreak, 
we can use ordinary message-passing equations for usual 
single-pathogen SIR~\cite{Karrer2010,Min2018}, that is
\begin{align}
v_{ij} &= 1- p \left( 1- \prod_{k \in \partial j \setminus i} v_{jk} \right).
\label{mps}
\end{align}

Then the probability of observing a coinfection outbreak initiated by two 
singly-infected seeds, $i$ and $j$, is
\begin{align}
P^{ij}_{ab} &=  \left( 1- \prod_{k \in \partial i} v_{ik} \right)
\left( 1- \prod_{l \in \partial j} v_{jl} \right).
\label{Pij}
\end{align}
The average probability for every pair of seeds $i$ and $j$ is then
\begin{align}
P_{ab} &= \frac{1}{\binom{N}{2}} \sum_{i,j} 
\left( 1- \prod_{k \in \partial i} v_{ik} \right)
\left( 1- \prod_{l \in \partial j} v_{jl} \right).
\label{P}
\end{align}
Notice that, since it is based on single-pathogen quantities, 
$P_{ab}$ changes continuously at the transition and has no 
dependence on $q$ whatsoever.

\section{On Random networks}

The equations presented in the previous section can be easily solved
numerically for any given network, thus providing predictions for all
observable of interest.  In order to obtain fully analytical
predictions it is possible to perform a further step, 
assuming that all variables $u_{ij}$ share the same value $u$
and the network structure is given by the degree distribution $P(k)$, 
so that Equation~(\ref{u}) can be written as
\begin{align}
1-u &= \sum_{k=1}^{\infty} \frac{k P(k)}{\langle k \rangle} \left[  \frac{1}{2} (p+q) \left( 1- u^{k-1} \right) \left( 1- u^k \right) \right. \nonumber \\
&\left . +p \left( 1- u^{k-1} \right)  u^k  \right] 
\label{uan}
\end{align}
In the same way the expressions for the outbreak size $\rho_{ab}$ and $\rho_a$
become
\begin{align}
\rho_{ab} &= \sum_k P(k) (1- u^k)^2. \\ \nonumber
\rho_{a} &= \sum_k P(k) (1- u^k)u^k,
\end{align}
while 
\begin{align}
P_{ab} &= \sum_k P(k) (1- v^k)^2.
\end{align}
where $v$ is the solution of Eq.~(\ref{uan}) for $q=p$.

Defining the function
\begin{align}
f(u) &= 1-u- \sum_{k=1}^{\infty} \frac{k P(k)}{\langle k \rangle}  \\
  &\times \left[  \frac{1}{2} (p+q) \left( 1- u^{k-1}- u^{k} +  u^{2 k-1} \right)
+p \left( u^{k} - u^{2k-1} \right) \right] \nonumber
\end{align}
the epidemic threshold is determined by the condition $f'(1)=0$, i.e., 
\begin{align}
p\sum_{k=1}^{\infty} \frac{k (k-1) P(k)}{\langle k \rangle} = 1.
\end{align}
Thus, the epidemic threshold is the same as the ordinary SIR model as
\begin{align}
p_c = \frac{\langle k \rangle}{\langle k^2 \rangle - \langle k \rangle}.
\label{pc}
\end{align}
The point where the transition becomes discontinuous can be identified by the
conditions $f'(1)=0$ and $f''(1)=0$.
Since
\begin{align}
f''(1) 
&= \frac{1}{\langle k \rangle} \left[ (2p-q) \langle k^3 \rangle - (4p-q)\langle k^2 \rangle + 2p \langle k \rangle \right]. 
\end{align}
the change from continuous to hybrid transition occurs for $w=q/p>w_c$ 
where
\begin{align}
w_c& = \frac{2 \langle k^3 \rangle - 4 \langle k^2 \rangle +2 \langle k \rangle}{\langle k^3 \rangle - \langle k^2 \rangle}.
\label{wc}
\end{align}

\begin{figure}
\includegraphics[width=0.9\columnwidth]{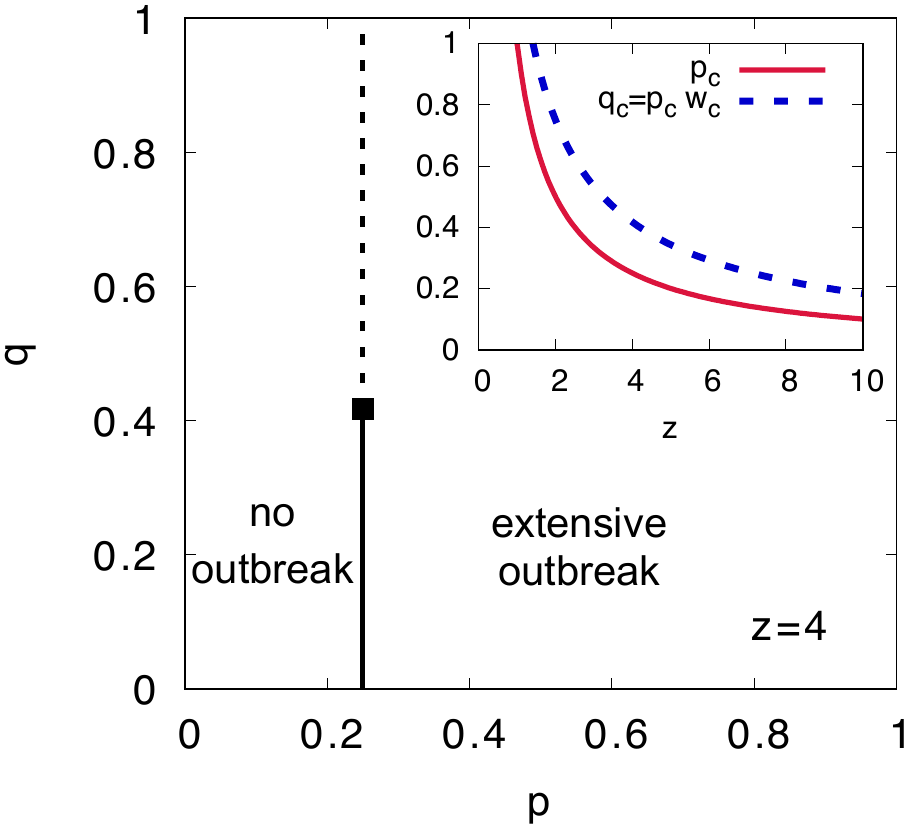}
\caption{
Phase diagram for ER networks obtained by the theory. 
The solid and dashed lines respectively represent 
continuous and discontinuous transitions.
The tricritical point $q_c$ where a continuous transition
becomes discontinuous is represented by a filled symbol.
(inset) Dependence on the average degree $z$ of the critical 
values $p_c$ and $q_c= p_c w_c$.
}
\label{pcqc}
\end{figure}

\subsection{Erd\H{o}s-R\'enyi networks}

For Erd\H{o}s-R\'enyi (ER) networks with average degree $\av{k}=z$, 
Eqs.~(\ref{pc}) and~(\ref{wc}) yield
\begin{align}
p_c &=\frac{1}{z}, \quad w_c =\frac{2(1+z)}{2+z}.
\label{thresholder}
\end{align}
In Fig.~\ref{pcqc}, we display, for fixed $z=4$, the phase-diagram of the model.
In the inset of Fig.~\ref{pcqc}, we plot the two critical values $p_c$ and $q_c=p_c w_c$
as a function of the average connectivity $z$.
While $p_c$ separates regions with and without extensive coinfection
outbreaks, $q_c = p_c w_c$ 
discriminates when the transition is continuous (for $q<q_c$)
and when it is discontinuous (for $q>q_c$).

\begin{figure*}
\includegraphics[width=\linewidth]{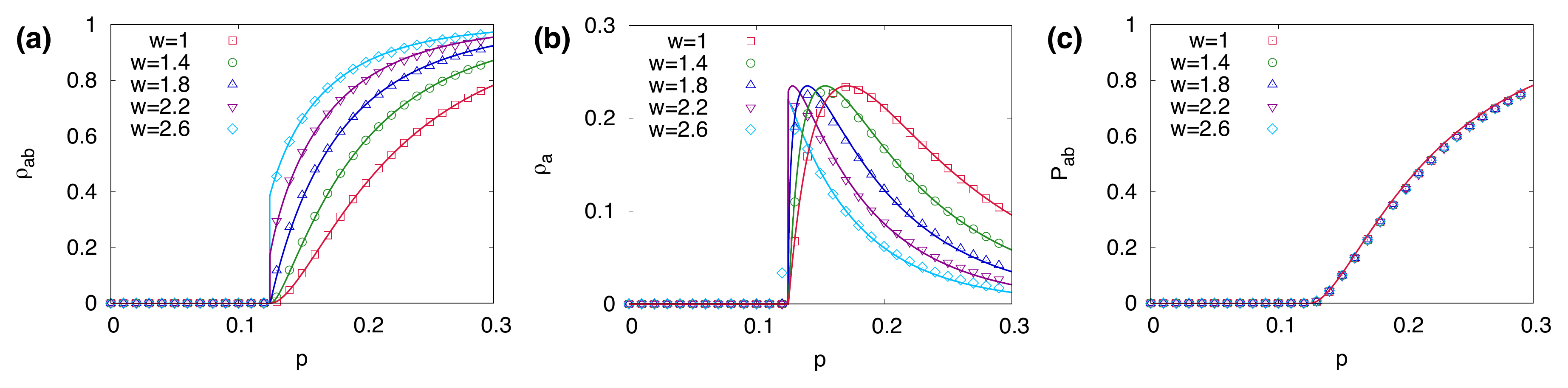}
\caption{
(a) Values of $\rho_{ab}$, (b) $\rho_{a}$ and of (c) the probability $P_{ab}$ 
of coinfection epidemic outbreaks on ER networks with $N=10^5$ and $\av{k}=8$.
For this choice of parameters, the theory predicts that $p_c=1/8$ and $w_c=1.8$.
Numerical results (symbols) are obtained by averaging over
$10^5$ realizations. 
Theoretical results (continuous lines) are obtained by means of 
Eqs.~(\ref{rhoaber}),~(\ref{rhoaer}) and~(\ref{Per}), respectively.
}
\label{fig:rhomp}
\end{figure*}

For ER networks we can determine analytically the behavior 
of relevant observables for any value of $p$ and $q$.
Poisson degree distribution implies that
$\sum_k P(k) u^{k} = {z}^{-1} \sum_k k P(k) u^{k-1}=e^{z(u-1)}$, 
so that the equation for the probability $u$ is
\begin{align}
1-u&=\frac{1}{2} (p+q) \left[ 1- e^{z(u-1)} - u e^{z(u-1)} + u e^{z(u^2-1)} \right] \nonumber \\
&+ p \left[ u e^{z(u-1)} - u e^{z(u^2-1)} \right]
\label{uer}
\end{align}
the equation for the outbreak size is
\begin{align}
\rho_{ab}&=1-2e^{z(u-1)}+e^{z(u^2-1)}, 
\label{rhoaber}
\end{align}
while 
\begin{align}
\rho_{a}&=e^{z(u-1)}-e^{z(u^2-1)},
\label{rhoaer}
\end{align}
and the probability to have a coinfection outbreak is
\begin{align}
P_{ab}&=1-2e^{z(v-1)}+e^{z(v^2-1)},
\label{Per}
\end{align}
where $v$ is the solution of
\begin{align}
1-v&=p \left[1 - e^{z(v-1)}\right].
\label{aer}
\end{align}

\subsection{Power-law distributed networks}

For power-law distributed networks with degree distribution
$P(k) = (\gamma-1) m^{\gamma-1} k^{-\gamma}$ (with $m$ the minimum degree)
the prediction for the threshold Eq.~(\ref{pc}) becomes
\begin{align}
p_c = \frac{1}{m \frac{\gamma-2}{\gamma-3}-1}.
\end{align}
Replacing the expressions for the moments of the distribution $P(k)$ in 
Eq.~(\ref{wc}) it turns out that the critical value $w_c$ increases as
$\gamma$ is lowered, up to a value $w_c=2$ for $\gamma=4$. For $\gamma<4$
the third moment of the degree distribution diverges and therefore, in
the infinite size limit Eq.~(\ref{wc}) predicts $w_c=2$.

\section{Numerical simulations}

We compare the analytical results obtained above with direct numerical 
simulations of the coinfective spreading dynamics. As contact
patterns for the epidemic process we consider either homogeneous 
ER networks or power-law distributed networks built according
to the configuration model \cite{Newmanbook}.
Numerically, we consider an outbreak to be epidemic if the final density
of coinfected nodes $\rho_{ab}$ is larger than a threshold equal to $0.001$.
We have checked that results do not change if different values are considered.

\subsection{Erd\H{o}s-R\'enyi networks}
In Fig.~\ref{fig:rhomp} we present a comparison between the theoretical
predictions of the message-passing approach, complemented with the 
random networks with a given degree distribution and numerical 
simulations performed on ER networks.
The agreement between theory and simulations is excellent overall.
In agreement with Eq.~(\ref{thresholder}),
the transition point $p_c$ does not depend on $w=q/p$.
The coinfection outbreak size $\rho_{ab}$ and the fraction $\rho_{a}$ 
of singly-infected nodes change continuously or discontinuously 
at the threshold depending on whether $w$ is smaller or larger 
than $w_c$. Conversely, the probability of epidemic outbreaks  
always undergoes a continuous transition and is completely independent
from $q$. 

\begin{figure}[b]
\includegraphics[width=\linewidth]{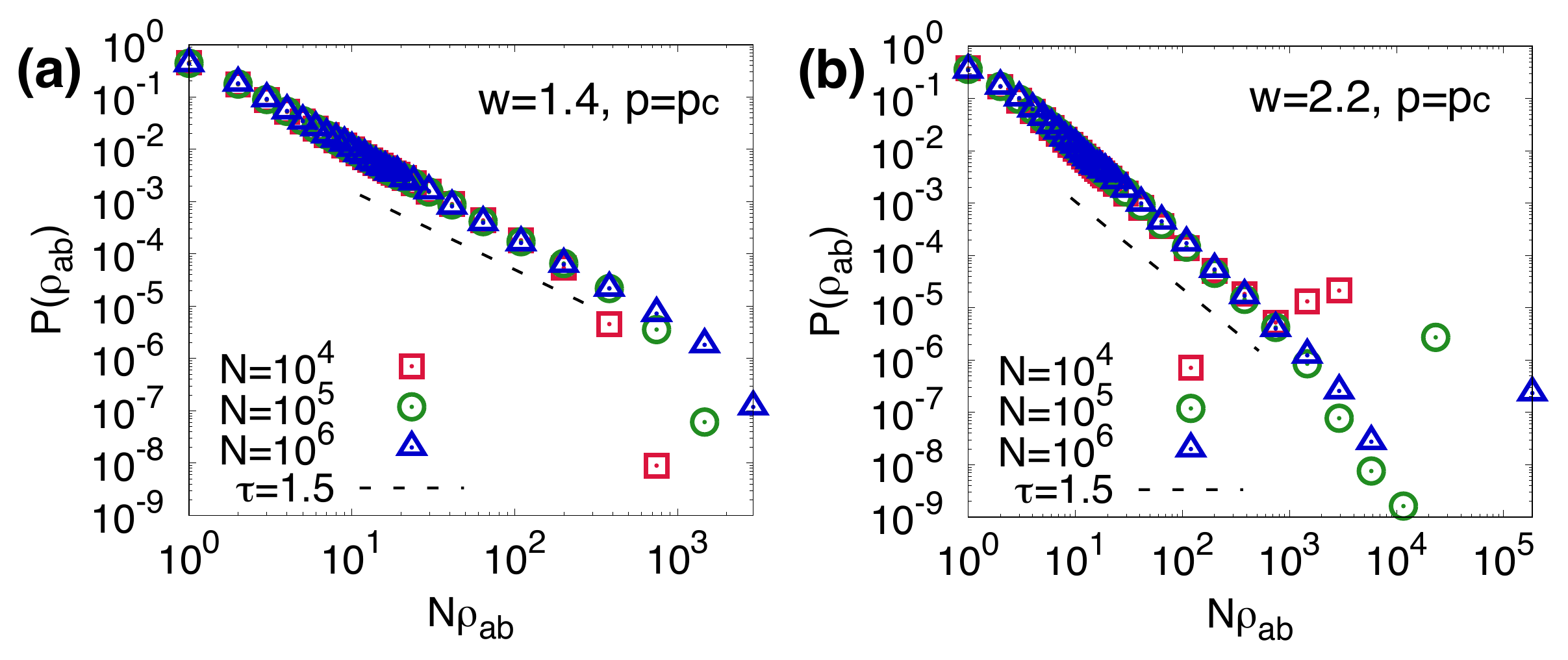}
\caption{
The size distribution of $\rho_{ab}$ at 
(a) $w=1.4$ and (b) $w=2.2$ and $p=p_c=0.125$
for ER networks with $N=10^4,\ 10^5,\ 10^6$ and $\langle k \rangle =8$.
}
\label{fig:rho3}
\end{figure}

Further insight into the nature of the transition
is provided in Fig.~\ref{fig:rho3}.
For different values of $w$ the distribution of avalanche sizes at the critical point
decays as a power-law with the same exponent $\tau=3/2$ valid for 
single-pathogen SIR dynamics.
Above the critical value $w_c$, however, the hybrid nature of the transition
results in the formation of a peak at a finite value of $\rho_{ab}$.

\begin{figure*}[t]
\includegraphics[width=\linewidth]{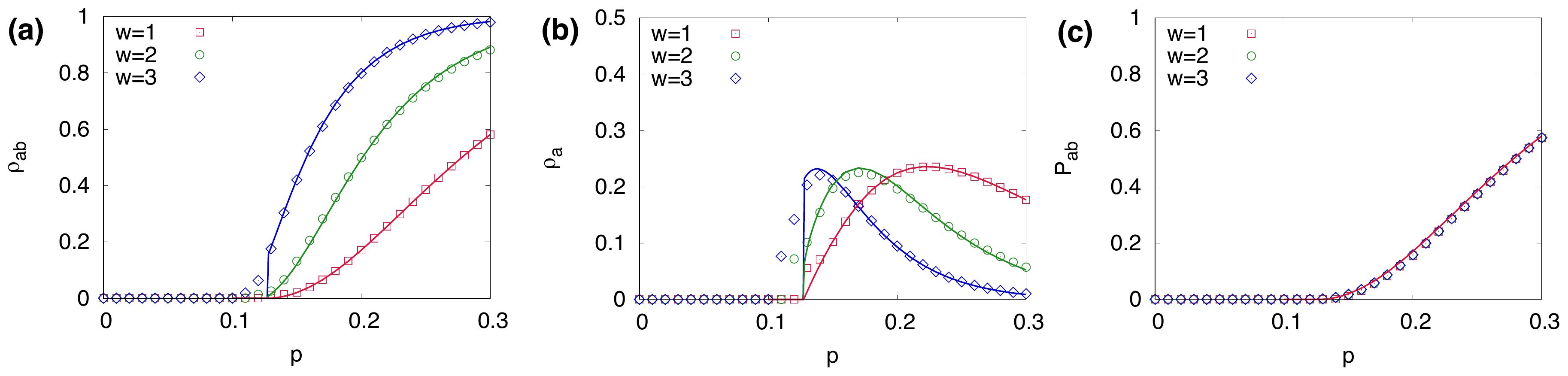}
\caption{
(a) Values of $\rho_{ab}$, (b) $\rho_{a}$ and (c) the probability $P_{ab}$ 
of coinfection epidemic outbreaks on a power-law distributed network
with $N=10^4$ and $\gamma=3.5$.
Numerical results (symbols) are obtained by averaging over
$10^5$ realizations. 
Theoretical results (continuous lines) are obtained by means of 
Eqs.~(\ref{rhoAB}),~(\ref{rhoA}) and~(\ref{P}), respectively.
}
\label{fig:rhomp2}
\end{figure*}

\subsection{Power-law distributed networks}

In Fig.~\ref{fig:rhomp2} we compare the theoretical
predictions of the message-passing approach and numerical results
obtained on power-law distributed networks. 
In this example, we generate a network 
with degree exponent $\gamma=3.5$ according to the 
configuration model \cite{Newmanbook}. 
The network with $\gamma=3.5$ is heterogeneous
but still maintains a non-zero epidemic threshold
in the thermodynamic limit. 
Also in this case the transition point depends only on 
$p$ and not on $q$, in agreement with Eq.~(\ref{threshold}).
What changes when $q$ increases is the nature of the transition.
Above a critical value $w=w_c$ for the ratio $w=q/p$ the
transition becomes hybrid.
The coinfection outbreak size $\rho_{ab}$ and $\rho_a$ both jump
discontinuously at the threshold, while the probability $P_{ab}$ of
epidemic outbreaks always undergoes a continuous transition and is 
completely independent from $q$.
Numerical results agree well with this picture.
There is a small discrepancy in $\rho_{ab}$ and $\rho_{a}$ below $p_c$
due to finite size effects. 
$\rho_{ab}$ and $\rho_{a}$ are average 
values calculated over realizations such that $\rho_{ab}>0.001$. 
Even if the probability of observing extensive outbreaks is zero below
the threshold in the thermodynamic limit, for finite system and 
sufficiently many realizations there is still a little, yet very low, chance
to reach the upper branch of the two stable solutions in the bistable region.
The values different from zero in Fig.~\ref{fig:rhomp2} immediately below
the threshold are the effect of these spurious events.

\begin{figure*}
\includegraphics[width=\linewidth]{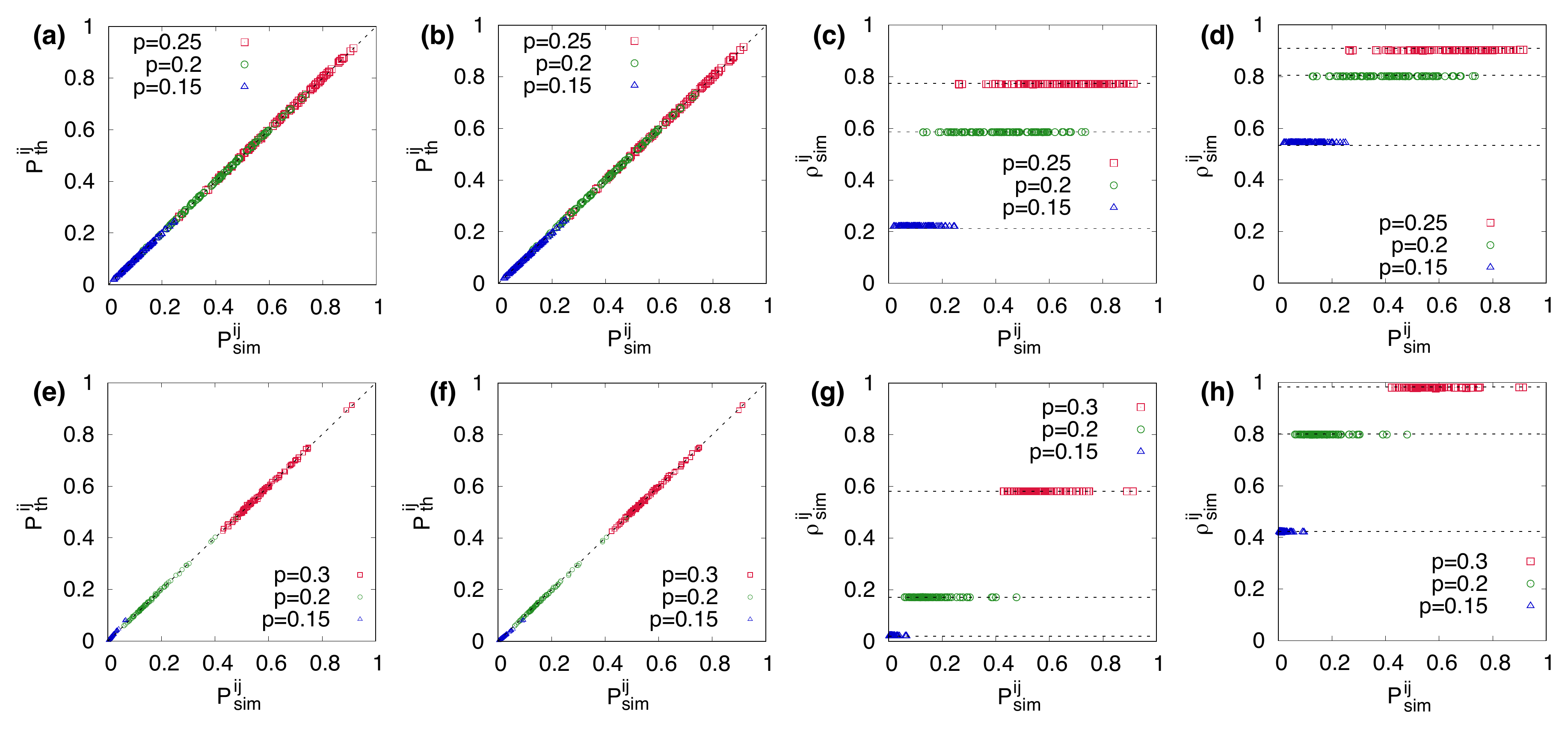}
\caption{
Comparison between the message-passing theory and numerical results for the 
probability of epidemic outbreaks for two singly-infected seeds 
when (a) $w=1.4$ and (b) $w=2.2$ on an ER network with $\av{k}=8$ and size $N=10^4$
and when (e) $w=1$ and (f) $w=3$ on a SF network with $\gamma=3.5$ and size $N=10^4$.
Comparison between the message-passing theory (dashed lines) and numerical 
results (symbols) for the size of epidemic outbreaks generated by two singly-infected 
seeds with (c) $w=1.4$ (c) and (d) $w=2.2$ on an ER network 
and with (g) $w=1$ and (h) $w=3$ on a SF network. 
}
\label{fig:influence}
\end{figure*}

\section{Finding influential spreaders}

We now turn to the identification of influential spreaders in the 
network, a problem which has attracted a large interest recently \cite{Kitsak2010,Lu2016}.
In practice we are interested in predicting, given
two specific nodes $i$ and $j$ as seeds
(each infected with a single-pathogen), what is the probability
that an extensive coinfection outbreak occurs and what is
its expected size.
Within the message-passing approach, 
the quantity $P^{ij}_{ab}$ calculated in Eq.~(\ref{Pij})
is the answer to the first question, while the
size of the outbreak is given by Eq.~(\ref{rhoAB}).

In Fig.~\ref{fig:influence} we compare
this analytical prediction to numerical results for an ER and a SF
network with the two initiators randomly chosen.
The message-passing approach allows us to determine with great
accuracy the ability of a pair of nodes to generate global coinfection
epidemic [see panels (a,b,e,f) of Fig.~\ref{fig:influence}].
Such ability depends strongly on $p$ and on
which pair of seeds is selected, while it is independent of the 
value of $q$, as predicted by Eq.~(\ref{Pij}).
Once an outbreak develops its size
is instead insensitive to the location of the initial seeds 
[see panels (c,d,g,h) of Fig.~\ref{fig:influence}],
while it changes depending on the value of the cooperativity
parameter $w$. This is again in excellent agreement with the 
theory. 

\begin{figure}
\includegraphics[width=\linewidth]{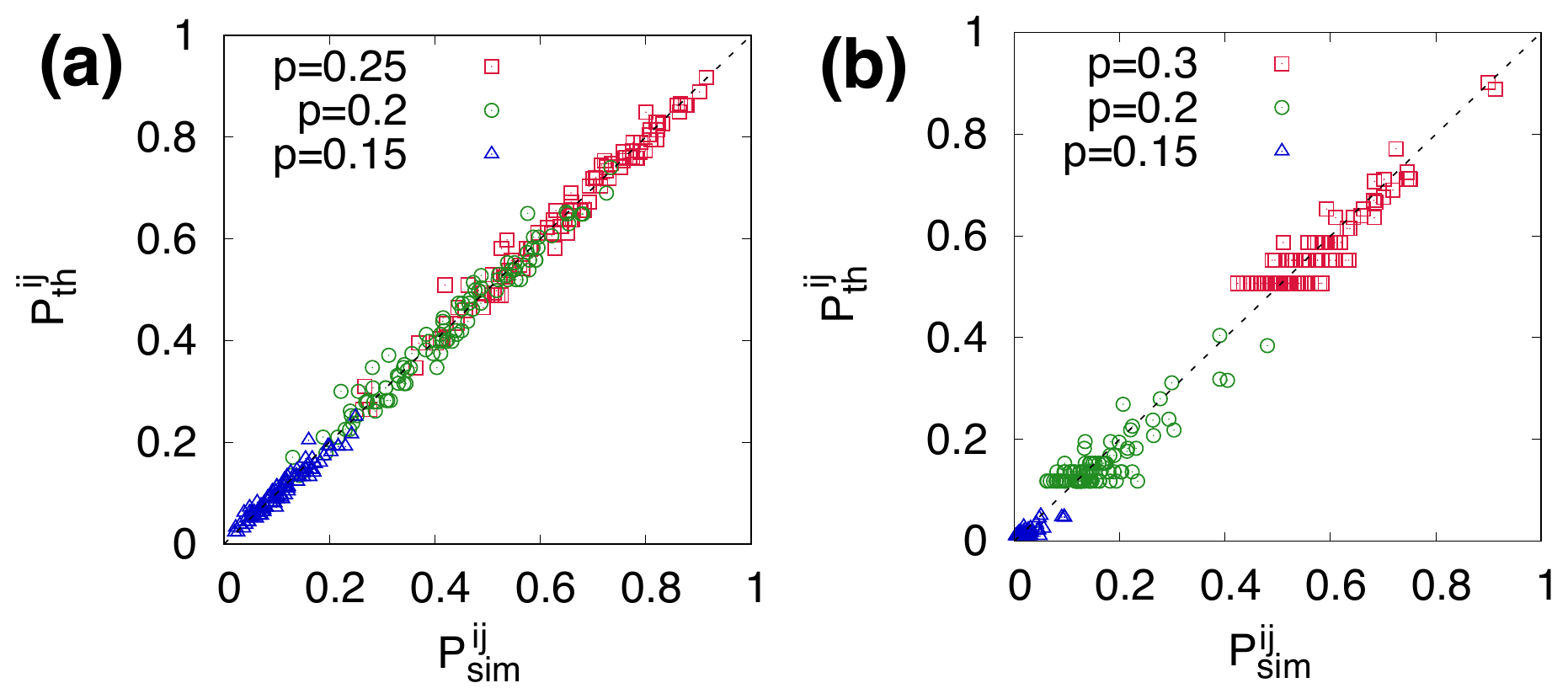}
\caption{
Comparison between the approximation (Eq.~(\ref{Pijan})) 
and numerical results for the probability of epidemic outbreaks for
two singly-infected seeds when (a) $w=2.2$ on an ER network 
with $\av{k}=8$ and size $N=10^4$
and (b) $w=3$ on a SF network with $\gamma=3.5$ and size $N=10^4$.
}
\label{fig:influence2}
\end{figure}

Eq.~(\ref{Pij}) provides an analytical expression
for the probability $P^{ij}_{ab}$ of observing an extensive
cooperative outbreak originated from nodes $i$ and $j$
\begin{align}
P^{ij}_{ab} = (1-v^{k_i})(1-v^{k_j}).
\label{Pijan}
\end{align}
In Fig.~\ref{fig:influence2} we test numerically this prediction finding
again a remarkably good agreement with relatively small computational
cost. 
In the vicinity of the critical point it is possible to set $v=1-\ep$ and
expand Eq.~(\ref{Pijan}) as a function of $\ep$, 
obtaining 
\begin{align}
P^{ij}_{ab} \sim \ep^2 k_i k_j.
\label{expansion}
\end{align}
This result connects the spreading influence of a pair of nodes with their
degree centrality. In the case a single doubly-infected seed is considered
Eq.~(\ref{expansion}) predicts a dependence on the square of the seed degree, 
in agreement with recent numerical results~\cite{Cui2019}.

\section{Conclusions}

In this paper, we propose a model of mutually cooperative coinfections 
with uncorrelated incoming transmission probabilities and study
the behavior of the model based on message-passing equations.
In this way we are able to provide
accurate predictions -- beyond the mean-field approximation -- 
of the phase-diagram, the probability of  epidemic outbreaks, 
the size of epidemic outbreaks for any given network structure. 
We confirm our predictions with numerical simulations on random networks
for both homogeneous and heterogeneous networks. 
By applying this theoretical framework we can calculate the spreading
influence of individual nodes and thus identify which of them
maximize the probability of a global extensive outbreak.
Our study provides a systematic way to analyze the cooperativity in 
spreading processes on complex networks.
Further studies will be needed to examine the case of nonsymmetric 
transmission probabilities, interacting epidemics in clustered 
networks (for which the present message-passing approach becomes 
inaccurate) or in meta-populations networks. 
Moreover, since most real-world contagion processes take place in 
multiple different layers of networks, a natural extension of our work
would be the consideration of cooperative epidemics on a multilayer 
network. While epidemic spreading on multilayer networks has been 
studied extensively~\cite{Mendiola2012,Kivela2014,Min2016}, 
cooperative epidemics on multilayer networks have not been fully
investigated. 
Our model and 
message-passing theory can guide to the study of cooperative epidemics
on multilayer networks.

\section{Acknowledgments}
We acknowledge financial support from the 2018 Short Term Mobility 
program of Consiglio Nazionale delle Ricerche (CNR) of Italy.
This work was supported by the National Research Foundation of
Korea (NRF) grant funded by the Korea government (MSIT) (No. 2018R1C1B5044202).
We also thank C. Han for high-performance computing support.

\bibliography{resub_min_castellano}

%merlin.mbs aipnum4-1.bst 2010-07-25 4.21a (PWD, AO, DPC) hacked
%Control: key (0)
%Control: author (8) initials jnrlst
%Control: editor formatted (1) identically to author
%Control: production of article title (0) allowed
%Control: page (1) range
%Control: year (1) truncated
%Control: production of eprint (0) enabled
\begin{thebibliography}{39}%
\makeatletter
\providecommand \@ifxundefined [1]{%
 \@ifx{#1\undefined}
}%
\providecommand \@ifnum [1]{%
 \ifnum #1\expandafter \@firstoftwo
 \else \expandafter \@secondoftwo
 \fi
}%
\providecommand \@ifx [1]{%
 \ifx #1\expandafter \@firstoftwo
 \else \expandafter \@secondoftwo
 \fi
}%
\providecommand \natexlab [1]{#1}%
\providecommand \enquote  [1]{``#1''}%
\providecommand \bibnamefont  [1]{#1}%
\providecommand \bibfnamefont [1]{#1}%
\providecommand \citenamefont [1]{#1}%
\providecommand \href@noop [0]{\@secondoftwo}%
\providecommand \href [0]{\begingroup \@sanitize@url \@href}%
\providecommand \@href[1]{\@@startlink{#1}\@@href}%
\providecommand \@@href[1]{\endgroup#1\@@endlink}%
\providecommand \@sanitize@url [0]{\catcode `\\12\catcode `\$12\catcode
  `\&12\catcode `\#12\catcode `\^12\catcode `\_12\catcode `\%12\relax}%
\providecommand \@@startlink[1]{}%
\providecommand \@@endlink[0]{}%
\providecommand \url  [0]{\begingroup\@sanitize@url \@url }%
\providecommand \@url [1]{\endgroup\@href {#1}{\urlprefix }}%
\providecommand \urlprefix  [0]{URL }%
\providecommand \Eprint [0]{\href }%
\providecommand \doibase [0]{http://dx.doi.org/}%
\providecommand \selectlanguage [0]{\@gobble}%
\providecommand \bibinfo  [0]{\@secondoftwo}%
\providecommand \bibfield  [0]{\@secondoftwo}%
\providecommand \translation [1]{[#1]}%
\providecommand \BibitemOpen [0]{}%
\providecommand \bibitemStop [0]{}%
\providecommand \bibitemNoStop [0]{.\EOS\space}%
\providecommand \EOS [0]{\spacefactor3000\relax}%
\providecommand \BibitemShut  [1]{\csname bibitem#1\endcsname}%
\let\auto@bib@innerbib\@empty
%</preamble>
\bibitem [{\citenamefont {Morens}, \citenamefont {Taubenberger},\ and\
  \citenamefont {Fauci}(2008)}]{Morens2008}%
  \BibitemOpen
  \bibfield  {author} {\bibinfo {author} {\bibfnamefont {D.~M.}\ \bibnamefont
  {Morens}}, \bibinfo {author} {\bibfnamefont {J.~K.}\ \bibnamefont
  {Taubenberger}}, \ and\ \bibinfo {author} {\bibfnamefont {A.~S.}\
  \bibnamefont {Fauci}},\ }\bibfield  {title} {\enquote {\bibinfo {title}
  {Predominant role of bacterial pneumonia as a cause of death in pandemic
  influenza: Implications for pandemic influenza preparedness},}\ }\href
  {\doibase 10.1086/591708} {\bibfield  {journal} {\bibinfo  {journal} {The
  Journal of Infectious Diseases}\ }\textbf {\bibinfo {volume} {198}},\
  \bibinfo {pages} {962} (\bibinfo {year} {2008})}\BibitemShut {NoStop}%
\bibitem [{\citenamefont {Brundage}\ and\ \citenamefont
  {Shanks}(2008)}]{Brundage2008}%
  \BibitemOpen
  \bibfield  {author} {\bibinfo {author} {\bibfnamefont {J.~F.}\ \bibnamefont
  {Brundage}}\ and\ \bibinfo {author} {\bibfnamefont {G.}~\bibnamefont
  {Shanks}},\ }\bibfield  {title} {\enquote {\bibinfo {title} {Deaths from
  bacterial pneumonia during 1918–19 influenza pandemic},}\ }\href {\doibase
  https://dx.doi.org/10.3201/eid1408.071313} {\bibfield  {journal} {\bibinfo
  {journal} {Emerg. Infect Dis.}\ }\textbf {\bibinfo {volume} {14}},\ \bibinfo
  {pages} {1193--1199} (\bibinfo {year} {2008})}\BibitemShut {NoStop}%
\bibitem [{\citenamefont {Sulkowski}(2008)}]{Sulkowski2008}%
  \BibitemOpen
  \bibfield  {author} {\bibinfo {author} {\bibfnamefont {M.~S.}\ \bibnamefont
  {Sulkowski}},\ }\bibfield  {title} {\enquote {\bibinfo {title} {{Viral
  hepatitis and HIV coinfection}},}\ }\href {\doibase
  http://dx.doi.org/10.1016/j.jhep.2007.11.009} {\bibfield  {journal} {\bibinfo
   {journal} {Journal of Hepatology}\ }\textbf {\bibinfo {volume} {48}},\
  \bibinfo {pages} {353--367} (\bibinfo {year} {2008})}\BibitemShut {NoStop}%
\bibitem [{\citenamefont {Pawlowski}\ \emph {et~al.}(2012)\citenamefont
  {Pawlowski}, \citenamefont {Jansseon}, \citenamefont {Sk\"old}, \citenamefont
  {Rottenberg},\ and\ \citenamefont {K\"allenius}}]{Pawlowski2012}%
  \BibitemOpen
  \bibfield  {author} {\bibinfo {author} {\bibfnamefont {A.}~\bibnamefont
  {Pawlowski}}, \bibinfo {author} {\bibfnamefont {M.}~\bibnamefont {Jansseon}},
  \bibinfo {author} {\bibfnamefont {M.}~\bibnamefont {Sk\"old}}, \bibinfo
  {author} {\bibfnamefont {M.~E.}\ \bibnamefont {Rottenberg}}, \ and\ \bibinfo
  {author} {\bibfnamefont {G.}~\bibnamefont {K\"allenius}},\ }\bibfield
  {title} {\enquote {\bibinfo {title} {Tuberculosis and hiv co-infection},}\
  }\href {\doibase 10.1371/journal.ppat.1002464} {\bibfield  {journal}
  {\bibinfo  {journal} {PloS Pathogens}\ }\textbf {\bibinfo {volume} {8(2)}},\
  \bibinfo {pages} {e1002464} (\bibinfo {year} {2012})}\BibitemShut {NoStop}%
\bibitem [{\citenamefont {Pastor-Satorras}\ \emph {et~al.}(2015)\citenamefont
  {Pastor-Satorras}, \citenamefont {Castellano}, \citenamefont {Van~Mieghem},\
  and\ \citenamefont {Vespignani}}]{PastorSatorras2015}%
  \BibitemOpen
  \bibfield  {author} {\bibinfo {author} {\bibfnamefont {R.}~\bibnamefont
  {Pastor-Satorras}}, \bibinfo {author} {\bibfnamefont {C.}~\bibnamefont
  {Castellano}}, \bibinfo {author} {\bibfnamefont {P.}~\bibnamefont
  {Van~Mieghem}}, \ and\ \bibinfo {author} {\bibfnamefont {A.}~\bibnamefont
  {Vespignani}},\ }\bibfield  {title} {\enquote {\bibinfo {title} {Epidemic
  processes in complex networks},}\ }\href {\doibase 10.1103/RevModPhys.87.925}
  {\bibfield  {journal} {\bibinfo  {journal} {Rev. Mod. Phys.}\ }\textbf
  {\bibinfo {volume} {87}},\ \bibinfo {pages} {925--979} (\bibinfo {year}
  {2015})}\BibitemShut {NoStop}%
\bibitem [{\citenamefont {Newman}(2010)}]{Newmanbook}%
  \BibitemOpen
  \bibfield  {author} {\bibinfo {author} {\bibfnamefont {M.}~\bibnamefont
  {Newman}},\ }\href@noop {} {\emph {\bibinfo {title} {Networks: An
  Introduction}}}\ (\bibinfo  {publisher} {Oxford University Press, Inc.},\
  \bibinfo {address} {New York, NY, USA},\ \bibinfo {year} {2010})\BibitemShut
  {NoStop}%
\bibitem [{\citenamefont {Dorogovtsev}, \citenamefont {Goltsev},\ and\
  \citenamefont {Mendes}(2008)}]{Dorogovtsev2008}%
  \BibitemOpen
  \bibfield  {author} {\bibinfo {author} {\bibfnamefont {S.~N.}\ \bibnamefont
  {Dorogovtsev}}, \bibinfo {author} {\bibfnamefont {A.~V.}\ \bibnamefont
  {Goltsev}}, \ and\ \bibinfo {author} {\bibfnamefont {J.~F.~F.}\ \bibnamefont
  {Mendes}},\ }\bibfield  {title} {\enquote {\bibinfo {title} {Critical
  phenomena in complex networks},}\ }\href {\doibase
  10.1103/RevModPhys.80.1275} {\bibfield  {journal} {\bibinfo  {journal} {Rev.
  Mod. Phys.}\ }\textbf {\bibinfo {volume} {80}},\ \bibinfo {pages}
  {1275--1335} (\bibinfo {year} {2008})}\BibitemShut {NoStop}%
\bibitem [{\citenamefont {Chen}\ \emph {et~al.}(2013)\citenamefont {Chen},
  \citenamefont {Ghanbarnejad}, \citenamefont {Cai},\ and\ \citenamefont
  {Grassberger}}]{Chen2013}%
  \BibitemOpen
  \bibfield  {author} {\bibinfo {author} {\bibfnamefont {L.}~\bibnamefont
  {Chen}}, \bibinfo {author} {\bibfnamefont {F.}~\bibnamefont {Ghanbarnejad}},
  \bibinfo {author} {\bibfnamefont {W.}~\bibnamefont {Cai}}, \ and\ \bibinfo
  {author} {\bibfnamefont {P.}~\bibnamefont {Grassberger}},\ }\bibfield
  {title} {\enquote {\bibinfo {title} {Outbreaks of coinfections: The critical
  role of cooperativity},}\ }\href {\doibase 10.1209/0295-5075/104/50001}
  {\bibfield  {journal} {\bibinfo  {journal} {EPL (Europhysics Letters)}\
  }\textbf {\bibinfo {volume} {104}},\ \bibinfo {pages} {50001} (\bibinfo
  {year} {2013})}\BibitemShut {NoStop}%
\bibitem [{\citenamefont {Cai}\ \emph {et~al.}(2015)\citenamefont {Cai},
  \citenamefont {Chen}, \citenamefont {Ghanbarnejad},\ and\ \citenamefont
  {Grassberger}}]{Chen2015}%
  \BibitemOpen
  \bibfield  {author} {\bibinfo {author} {\bibfnamefont {W.}~\bibnamefont
  {Cai}}, \bibinfo {author} {\bibfnamefont {L.}~\bibnamefont {Chen}}, \bibinfo
  {author} {\bibfnamefont {F.}~\bibnamefont {Ghanbarnejad}}, \ and\ \bibinfo
  {author} {\bibfnamefont {P.}~\bibnamefont {Grassberger}},\ }\bibfield
  {title} {\enquote {\bibinfo {title} {Avalanche outbreaks emerging in
  cooperative contagions},}\ }\href {\doibase 10.1038/nphys3457} {\bibfield
  {journal} {\bibinfo  {journal} {Nature physics}\ }\textbf {\bibinfo {volume}
  {11}},\ \bibinfo {pages} {936--940} (\bibinfo {year} {2015})}\BibitemShut
  {NoStop}%
\bibitem [{\citenamefont {Grassberger}\ \emph {et~al.}(2016)\citenamefont
  {Grassberger}, \citenamefont {Chen}, \citenamefont {Ghanbarnejad},\ and\
  \citenamefont {Cai}}]{Chen2016}%
  \BibitemOpen
  \bibfield  {author} {\bibinfo {author} {\bibfnamefont {P.}~\bibnamefont
  {Grassberger}}, \bibinfo {author} {\bibfnamefont {L.}~\bibnamefont {Chen}},
  \bibinfo {author} {\bibfnamefont {F.}~\bibnamefont {Ghanbarnejad}}, \ and\
  \bibinfo {author} {\bibfnamefont {W.}~\bibnamefont {Cai}},\ }\bibfield
  {title} {\enquote {\bibinfo {title} {Phase transitions in cooperative
  coinfections: Simulation results for networks and lattices},}\ }\href
  {\doibase 10.1103/PhysRevE.93.042316} {\bibfield  {journal} {\bibinfo
  {journal} {Phys. Rev. E}\ }\textbf {\bibinfo {volume} {93}},\ \bibinfo
  {pages} {042316} (\bibinfo {year} {2016})}\BibitemShut {NoStop}%
\bibitem [{\citenamefont {Janssen}\ and\ \citenamefont
  {Stenull}(2016)}]{Janssen2016}%
  \BibitemOpen
  \bibfield  {author} {\bibinfo {author} {\bibfnamefont {H.-K.}\ \bibnamefont
  {Janssen}}\ and\ \bibinfo {author} {\bibfnamefont {O.}~\bibnamefont
  {Stenull}},\ }\bibfield  {title} {\enquote {\bibinfo {title} {First-order
  phase transitions in outbreaks of co-infectious diseases and the extended
  general epidemic process},}\ }\href {\doibase 10.1209/0295-5075/113/26005}
  {\bibfield  {journal} {\bibinfo  {journal} {EPL (Europhys. Lett.)}\ }\textbf
  {\bibinfo {volume} {113}} (\bibinfo {year} {2016}),\
  10.1209/0295-5075/113/26005}\BibitemShut {NoStop}%
\bibitem [{\citenamefont {Choi}, \citenamefont {Lee},\ and\ \citenamefont
  {Kahng}(2017)}]{Kahngpre2017}%
  \BibitemOpen
  \bibfield  {author} {\bibinfo {author} {\bibfnamefont {W.}~\bibnamefont
  {Choi}}, \bibinfo {author} {\bibfnamefont {D.}~\bibnamefont {Lee}}, \ and\
  \bibinfo {author} {\bibfnamefont {B.}~\bibnamefont {Kahng}},\ }\bibfield
  {title} {\enquote {\bibinfo {title} {Mixed-order phase transition in a
  two-step contagion model with a single infectious seed},}\ }\href {\doibase
  10.1103/PhysRevE.95.022304} {\bibfield  {journal} {\bibinfo  {journal} {Phys.
  Rev. E}\ }\textbf {\bibinfo {volume} {95}} (\bibinfo {year} {2017}),\
  10.1103/PhysRevE.95.022304}\BibitemShut {NoStop}%
\bibitem [{\citenamefont {Goltsev}, \citenamefont {Dorogovtsev},\ and\
  \citenamefont {Mendes}(2006)}]{Goltsev2006}%
  \BibitemOpen
  \bibfield  {author} {\bibinfo {author} {\bibfnamefont {A.~V.}\ \bibnamefont
  {Goltsev}}, \bibinfo {author} {\bibfnamefont {S.~N.}\ \bibnamefont
  {Dorogovtsev}}, \ and\ \bibinfo {author} {\bibfnamefont {J.~F.~F.}\
  \bibnamefont {Mendes}},\ }\bibfield  {title} {\enquote {\bibinfo {title}
  {$k$},}\ }\href {\doibase 10.1103/PhysRevE.73.056101} {\bibfield  {journal}
  {\bibinfo  {journal} {Phys. Rev. E}\ }\textbf {\bibinfo {volume} {73}},\
  \bibinfo {pages} {056101} (\bibinfo {year} {2006})}\BibitemShut {NoStop}%
\bibitem [{\citenamefont {Min}\ and\ \citenamefont
  {San~Miguel}(2018)}]{MinMaxi2018}%
  \BibitemOpen
  \bibfield  {author} {\bibinfo {author} {\bibfnamefont {B.}~\bibnamefont
  {Min}}\ and\ \bibinfo {author} {\bibfnamefont {M.}~\bibnamefont
  {San~Miguel}},\ }\bibfield  {title} {\enquote {\bibinfo {title} {Competing
  contagion processes: complex contagion triggered by simple contagion},}\
  }\href {\doibase 10.1038/s41598-018-28615-3} {\bibfield  {journal} {\bibinfo
  {journal} {Scientific Reports}\ }\textbf {\bibinfo {volume} {8}},\ \bibinfo
  {pages} {10422} (\bibinfo {year} {2018})}\BibitemShut {NoStop}%
\bibitem [{\citenamefont {Lee}\ \emph {et~al.}(2017)\citenamefont {Lee},
  \citenamefont {Choi}, \citenamefont {Kert\'esz},\ and\ \citenamefont
  {Kahng}}]{Kahng2017}%
  \BibitemOpen
  \bibfield  {author} {\bibinfo {author} {\bibfnamefont {D.}~\bibnamefont
  {Lee}}, \bibinfo {author} {\bibfnamefont {W.}~\bibnamefont {Choi}}, \bibinfo
  {author} {\bibfnamefont {J.}~\bibnamefont {Kert\'esz}}, \ and\ \bibinfo
  {author} {\bibfnamefont {B.}~\bibnamefont {Kahng}},\ }\bibfield  {title}
  {\enquote {\bibinfo {title} {Universal mechanism for hybrid percolation
  transitions},}\ }\href {\doibase 10.1038/s41598-017-06182-3} {\bibfield
  {journal} {\bibinfo  {journal} {Scientific Reports}\ }\textbf {\bibinfo
  {volume} {7}} (\bibinfo {year} {2017}),\
  10.1038/s41598-017-06182-3}\BibitemShut {NoStop}%
\bibitem [{\citenamefont {Baek}\ \emph {et~al.}(2019)\citenamefont {Baek},
  \citenamefont {Chung}, \citenamefont {Ha}, \citenamefont {Jeong},\ and\
  \citenamefont {Kim}}]{Baek2019}%
  \BibitemOpen
  \bibfield  {author} {\bibinfo {author} {\bibfnamefont {Y.}~\bibnamefont
  {Baek}}, \bibinfo {author} {\bibfnamefont {K.}~\bibnamefont {Chung}},
  \bibinfo {author} {\bibfnamefont {M.}~\bibnamefont {Ha}}, \bibinfo {author}
  {\bibfnamefont {H.}~\bibnamefont {Jeong}}, \ and\ \bibinfo {author}
  {\bibfnamefont {D.}~\bibnamefont {Kim}},\ }\bibfield  {title} {\enquote
  {\bibinfo {title} {Role of hubs in the synergistic spread of behavior},}\
  }\href {\doibase 10.1103/PhysRevE.99.020301} {\bibfield  {journal} {\bibinfo
  {journal} {Phys. Rev. E}\ }\textbf {\bibinfo {volume} {99}},\ \bibinfo
  {pages} {020301} (\bibinfo {year} {2019})}\BibitemShut {NoStop}%
\bibitem [{\citenamefont {Cui}, \citenamefont {Colaiori},\ and\ \citenamefont
  {Castellano}(2017)}]{Cui2017}%
  \BibitemOpen
  \bibfield  {author} {\bibinfo {author} {\bibfnamefont {P.-B.}\ \bibnamefont
  {Cui}}, \bibinfo {author} {\bibfnamefont {F.}~\bibnamefont {Colaiori}}, \
  and\ \bibinfo {author} {\bibfnamefont {C.}~\bibnamefont {Castellano}},\
  }\bibfield  {title} {\enquote {\bibinfo {title} {Mutually cooperative
  epidemics on power-law networks},}\ }\href {\doibase
  10.1103/PhysRevE.96.022301} {\bibfield  {journal} {\bibinfo  {journal} {Phys.
  Rev. E}\ }\textbf {\bibinfo {volume} {96}},\ \bibinfo {pages} {022301}
  (\bibinfo {year} {2017})}\BibitemShut {NoStop}%
\bibitem [{\citenamefont {H{\'e}bert-Dufresne}\ and\ \citenamefont
  {Althouse}(2015)}]{HebertDufresne2015}%
  \BibitemOpen
  \bibfield  {author} {\bibinfo {author} {\bibfnamefont {L.}~\bibnamefont
  {H{\'e}bert-Dufresne}}\ and\ \bibinfo {author} {\bibfnamefont {B.~M.}\
  \bibnamefont {Althouse}},\ }\bibfield  {title} {\enquote {\bibinfo {title}
  {Complex dynamics of synergistic coinfections on realistically clustered
  networks},}\ }\href {\doibase 10.1073/pnas.1507820112} {\bibfield  {journal}
  {\bibinfo  {journal} {Proceedings of the National Academy of Sciences}\
  }\textbf {\bibinfo {volume} {112}},\ \bibinfo {pages} {10551--10556}
  (\bibinfo {year} {2015})},\ \Eprint
  {http://arxiv.org/abs/https://www.pnas.org/content/112/33/10551.full.pdf}
  {https://www.pnas.org/content/112/33/10551.full.pdf} \BibitemShut {NoStop}%
\bibitem [{\citenamefont {Cui}, \citenamefont {Colaiori},\ and\ \citenamefont
  {Castellano}(2019)}]{Cui2019}%
  \BibitemOpen
  \bibfield  {author} {\bibinfo {author} {\bibfnamefont {P.-B.}\ \bibnamefont
  {Cui}}, \bibinfo {author} {\bibfnamefont {F.}~\bibnamefont {Colaiori}}, \
  and\ \bibinfo {author} {\bibfnamefont {C.}~\bibnamefont {Castellano}},\
  }\bibfield  {title} {\enquote {\bibinfo {title} {Effect of network clustering
  on mutually cooperative coinfections},}\ }\href {\doibase
  10.1103/PhysRevE.99.022301} {\bibfield  {journal} {\bibinfo  {journal} {Phys.
  Rev. E}\ }\textbf {\bibinfo {volume} {99}},\ \bibinfo {pages} {022301}
  (\bibinfo {year} {2019})}\BibitemShut {NoStop}%
\bibitem [{\citenamefont {Azimi-Tafreshi}(2016)}]{AzimiTafreshi2016}%
  \BibitemOpen
  \bibfield  {author} {\bibinfo {author} {\bibfnamefont {N.}~\bibnamefont
  {Azimi-Tafreshi}},\ }\bibfield  {title} {\enquote {\bibinfo {title}
  {Cooperative epidemics on multiplex networks},}\ }\href {\doibase
  10.1103/PhysRevE.93.042303} {\bibfield  {journal} {\bibinfo  {journal} {Phys.
  Rev. E}\ }\textbf {\bibinfo {volume} {93}},\ \bibinfo {pages} {042303}
  (\bibinfo {year} {2016})}\BibitemShut {NoStop}%
\bibitem [{\citenamefont {Wei}\ \emph {et~al.}(2016)\citenamefont {Wei},
  \citenamefont {Chen}, \citenamefont {Wu}, \citenamefont {Feng},\ and\
  \citenamefont {an~Lu}}]{Wei2016}%
  \BibitemOpen
  \bibfield  {author} {\bibinfo {author} {\bibfnamefont {X.}~\bibnamefont
  {Wei}}, \bibinfo {author} {\bibfnamefont {S.}~\bibnamefont {Chen}}, \bibinfo
  {author} {\bibfnamefont {X.}~\bibnamefont {Wu}}, \bibinfo {author}
  {\bibfnamefont {J.}~\bibnamefont {Feng}}, \ and\ \bibinfo {author}
  {\bibfnamefont {J.}~\bibnamefont {an~Lu}},\ }\bibfield  {title} {\enquote
  {\bibinfo {title} {A unified framework of interplay between two spreading
  processes in multiplex networks},}\ }\href {\doibase
  10.1209/0295-5075/114/26006} {\bibfield  {journal} {\bibinfo  {journal}
  {{EPL} (Europhysics Letters)}\ }\textbf {\bibinfo {volume} {114}},\ \bibinfo
  {pages} {26006} (\bibinfo {year} {2016})}\BibitemShut {NoStop}%
\bibitem [{\citenamefont {Liu}\ \emph {et~al.}(2018)\citenamefont {Liu},
  \citenamefont {Wang}, \citenamefont {Cai}, \citenamefont {Tang},\ and\
  \citenamefont {Lai}}]{Liu2018}%
  \BibitemOpen
  \bibfield  {author} {\bibinfo {author} {\bibfnamefont {Q.-H.}\ \bibnamefont
  {Liu}}, \bibinfo {author} {\bibfnamefont {W.}~\bibnamefont {Wang}}, \bibinfo
  {author} {\bibfnamefont {S.-M.}\ \bibnamefont {Cai}}, \bibinfo {author}
  {\bibfnamefont {M.}~\bibnamefont {Tang}}, \ and\ \bibinfo {author}
  {\bibfnamefont {Y.-C.}\ \bibnamefont {Lai}},\ }\bibfield  {title} {\enquote
  {\bibinfo {title} {Synergistic interactions promote behavior spreading and
  alter phase transitions on multiplex networks},}\ }\href {\doibase
  10.1103/PhysRevE.97.022311} {\bibfield  {journal} {\bibinfo  {journal} {Phys.
  Rev. E}\ }\textbf {\bibinfo {volume} {97}},\ \bibinfo {pages} {022311}
  (\bibinfo {year} {2018})}\BibitemShut {NoStop}%
\bibitem [{\citenamefont {Rodríguez}, \citenamefont {Ghanbarnejad},\ and\
  \citenamefont {Eguíluz}(2017)}]{Rodriguez2017}%
  \BibitemOpen
  \bibfield  {author} {\bibinfo {author} {\bibfnamefont {J.~P.}\ \bibnamefont
  {Rodríguez}}, \bibinfo {author} {\bibfnamefont {F.}~\bibnamefont
  {Ghanbarnejad}}, \ and\ \bibinfo {author} {\bibfnamefont {V.~M.}\
  \bibnamefont {Eguíluz}},\ }\bibfield  {title} {\enquote {\bibinfo {title}
  {Risk of coinfection outbreaks in temporal networks: A case study of a
  hospital contact network},}\ }\href {\doibase 10.3389/fphy.2017.00046}
  {\bibfield  {journal} {\bibinfo  {journal} {Frontiers in Physics}\ }\textbf
  {\bibinfo {volume} {5}},\ \bibinfo {pages} {46} (\bibinfo {year}
  {2017})}\BibitemShut {NoStop}%
\bibitem [{\citenamefont {Chen}, \citenamefont {Ghanbarnejad},\ and\
  \citenamefont {Brockmann}(2017)}]{Chen2017}%
  \BibitemOpen
  \bibfield  {author} {\bibinfo {author} {\bibfnamefont {L.}~\bibnamefont
  {Chen}}, \bibinfo {author} {\bibfnamefont {F.}~\bibnamefont {Ghanbarnejad}},
  \ and\ \bibinfo {author} {\bibfnamefont {D.}~\bibnamefont {Brockmann}},\
  }\bibfield  {title} {\enquote {\bibinfo {title} {Fundamental properties of
  cooperative contagion processes},}\ }\href
  {http://stacks.iop.org/1367-2630/19/i=10/a=103041} {\bibfield  {journal}
  {\bibinfo  {journal} {New Journal of Physics}\ }\textbf {\bibinfo {volume}
  {19}},\ \bibinfo {pages} {103041} (\bibinfo {year} {2017})}\BibitemShut
  {NoStop}%
\bibitem [{\citenamefont {Zarei}, \citenamefont {Moghimi-Araghi},\ and\
  \citenamefont {Ghanbarnejad}(2019)}]{Zarei2019}%
  \BibitemOpen
  \bibfield  {author} {\bibinfo {author} {\bibfnamefont {F.}~\bibnamefont
  {Zarei}}, \bibinfo {author} {\bibfnamefont {S.}~\bibnamefont
  {Moghimi-Araghi}}, \ and\ \bibinfo {author} {\bibfnamefont {F.}~\bibnamefont
  {Ghanbarnejad}},\ }\bibfield  {title} {\enquote {\bibinfo {title} {Exact
  solution of generalized cooperative susceptible-infected-removed (sir)
  dynamics},}\ }\href {\doibase 10.1103/PhysRevE.100.012307} {\bibfield
  {journal} {\bibinfo  {journal} {Phys. Rev. E}\ }\textbf {\bibinfo {volume}
  {100}},\ \bibinfo {pages} {012307} (\bibinfo {year} {2019})}\BibitemShut
  {NoStop}%
\bibitem [{\citenamefont {Karrer}\ and\ \citenamefont
  {Newman}(2010)}]{Karrer2010}%
  \BibitemOpen
  \bibfield  {author} {\bibinfo {author} {\bibfnamefont {B.}~\bibnamefont
  {Karrer}}\ and\ \bibinfo {author} {\bibfnamefont {M.~E.~J.}\ \bibnamefont
  {Newman}},\ }\bibfield  {title} {\enquote {\bibinfo {title} {Message passing
  approach for general epidemic models},}\ }\href {\doibase
  10.1103/PhysRevE.82.016101} {\bibfield  {journal} {\bibinfo  {journal} {Phys.
  Rev. E}\ }\textbf {\bibinfo {volume} {82}},\ \bibinfo {pages} {016101}
  (\bibinfo {year} {2010})}\BibitemShut {NoStop}%
\bibitem [{\citenamefont {Lokhov}\ \emph {et~al.}(2014)\citenamefont {Lokhov},
  \citenamefont {M\'ezard}, \citenamefont {Ohta},\ and\ \citenamefont
  {Zdeborov\'a}}]{Lokhov2014}%
  \BibitemOpen
  \bibfield  {author} {\bibinfo {author} {\bibfnamefont {A.~Y.}\ \bibnamefont
  {Lokhov}}, \bibinfo {author} {\bibfnamefont {M.}~\bibnamefont {M\'ezard}},
  \bibinfo {author} {\bibfnamefont {H.}~\bibnamefont {Ohta}}, \ and\ \bibinfo
  {author} {\bibfnamefont {L.}~\bibnamefont {Zdeborov\'a}},\ }\bibfield
  {title} {\enquote {\bibinfo {title} {Inferring the origin of an epidemic with
  a dynamic message-passing algorithm},}\ }\href {\doibase
  10.1103/PhysRevE.90.012801} {\bibfield  {journal} {\bibinfo  {journal} {Phys.
  Rev. E}\ }\textbf {\bibinfo {volume} {90}},\ \bibinfo {pages} {012801}
  (\bibinfo {year} {2014})}\BibitemShut {NoStop}%
\bibitem [{\citenamefont {Min}(2018)}]{Min2018}%
  \BibitemOpen
  \bibfield  {author} {\bibinfo {author} {\bibfnamefont {B.}~\bibnamefont
  {Min}},\ }\bibfield  {title} {\enquote {\bibinfo {title} {Identifying an
  influential spreader from a single seed in complex network s via a
  message-passing approach},}\ }\href {\doibase 10.1140/epjb/e2017-80597-1}
  {\bibfield  {journal} {\bibinfo  {journal} {The European Physical Journal B}\
  }\textbf {\bibinfo {volume} {91}},\ \bibinfo {pages} {18} (\bibinfo {year}
  {2018})}\BibitemShut {NoStop}%
\bibitem [{\citenamefont {Kitsak}\ \emph {et~al.}(2010)\citenamefont {Kitsak},
  \citenamefont {Gallos}, \citenamefont {Havlin}, \citenamefont {L~iljeros},
  \citenamefont {Muchnik}, \citenamefont {Stanley},\ and\ \citenamefont
  {Makse}}]{Kitsak2010}%
  \BibitemOpen
  \bibfield  {author} {\bibinfo {author} {\bibfnamefont {M.}~\bibnamefont
  {Kitsak}}, \bibinfo {author} {\bibfnamefont {L.~K.}\ \bibnamefont {Gallos}},
  \bibinfo {author} {\bibfnamefont {S.}~\bibnamefont {Havlin}}, \bibinfo
  {author} {\bibfnamefont {F.}~\bibnamefont {L~iljeros}}, \bibinfo {author}
  {\bibfnamefont {L.}~\bibnamefont {Muchnik}}, \bibinfo {author} {\bibfnamefont
  {H.~E.}\ \bibnamefont {Stanley}}, \ and\ \bibinfo {author} {\bibfnamefont
  {H.~A.}\ \bibnamefont {Makse}},\ }\bibfield  {title} {\enquote {\bibinfo
  {title} {Identification of influential spreaders in complex networks},}\
  }\href {\doibase https://doi.org/10.1038/nphys1746} {\bibfield  {journal}
  {\bibinfo  {journal} {Nature Physics}\ }\textbf {\bibinfo {volume} {6}},\
  \bibinfo {pages} {888--893} (\bibinfo {year} {2010})}\BibitemShut {NoStop}%
\bibitem [{\citenamefont {Lu}\ \emph {et~al.}(2016)\citenamefont {Lu},
  \citenamefont {Chen}, \citenamefont {Ren}, \citenamefont {Zhang},
  \citenamefont {Zhang},\ and\ \citenamefont {Zhou}}]{Lu2016}%
  \BibitemOpen
  \bibfield  {author} {\bibinfo {author} {\bibfnamefont {L.}~\bibnamefont
  {Lu}}, \bibinfo {author} {\bibfnamefont {D.}~\bibnamefont {Chen}}, \bibinfo
  {author} {\bibfnamefont {X.-L.}\ \bibnamefont {Ren}}, \bibinfo {author}
  {\bibfnamefont {Q.-M.}\ \bibnamefont {Zhang}}, \bibinfo {author}
  {\bibfnamefont {Y.-C.}\ \bibnamefont {Zhang}}, \ and\ \bibinfo {author}
  {\bibfnamefont {T.}~\bibnamefont {Zhou}},\ }\bibfield  {title} {\enquote
  {\bibinfo {title} {Vital nodes identification in complex networks},}\ }\href
  {\doibase https://doi.org/10.1016/j.physrep.2016.06.007} {\bibfield
  {journal} {\bibinfo  {journal} {Physics Reports}\ }\textbf {\bibinfo {volume}
  {650}},\ \bibinfo {pages} {1 -- 63} (\bibinfo {year} {2016})}\BibitemShut
  {NoStop}%
\bibitem [{\citenamefont {Radicchi}\ and\ \citenamefont
  {Castellano}(2016{\natexlab{a}})}]{Radicchi2016}%
  \BibitemOpen
  \bibfield  {author} {\bibinfo {author} {\bibfnamefont {F.}~\bibnamefont
  {Radicchi}}\ and\ \bibinfo {author} {\bibfnamefont {C.}~\bibnamefont
  {Castellano}},\ }\bibfield  {title} {\enquote {\bibinfo {title} {Leveraging
  percolation theory to single out influential spreaders in networks},}\ }\href
  {\doibase 10.1103/PhysRevE.93.062314} {\bibfield  {journal} {\bibinfo
  {journal} {Phys. Rev. E}\ }\textbf {\bibinfo {volume} {93}},\ \bibinfo
  {pages} {062314} (\bibinfo {year} {2016}{\natexlab{a}})}\BibitemShut
  {NoStop}%
\bibitem [{\citenamefont {Newman}(2009)}]{Newman2009}%
  \BibitemOpen
  \bibfield  {author} {\bibinfo {author} {\bibfnamefont {M.~E.~J.}\
  \bibnamefont {Newman}},\ }\bibfield  {title} {\enquote {\bibinfo {title}
  {Random graphs with clustering},}\ }\href {\doibase
  10.1103/PhysRevLett.103.058701} {\bibfield  {journal} {\bibinfo  {journal}
  {Phys. Rev. Lett.}\ }\textbf {\bibinfo {volume} {103}} (\bibinfo {year}
  {2009}),\ 10.1103/PhysRevLett.103.058701}\BibitemShut {NoStop}%
\bibitem [{\citenamefont {Radicchi}\ and\ \citenamefont
  {Castellano}(2016{\natexlab{b}})}]{Radicchiprer2016}%
  \BibitemOpen
  \bibfield  {author} {\bibinfo {author} {\bibfnamefont {F.}~\bibnamefont
  {Radicchi}}\ and\ \bibinfo {author} {\bibfnamefont {C.}~\bibnamefont
  {Castellano}},\ }\bibfield  {title} {\enquote {\bibinfo {title} {Beyond the
  locally treelike approximation for percolation on real networks},}\ }\href
  {\doibase 10.1103/PhysRevE.93.030302} {\bibfield  {journal} {\bibinfo
  {journal} {Phys. Rev. E}\ }\textbf {\bibinfo {volume} {93}} (\bibinfo {year}
  {2016}{\natexlab{b}}),\ 10.1103/PhysRevE.93.030302}\BibitemShut {NoStop}%
\bibitem [{\citenamefont {Cantwell}\ and\ \citenamefont
  {Newman}(2019)}]{Cantwell2019}%
  \BibitemOpen
  \bibfield  {author} {\bibinfo {author} {\bibfnamefont {G.~T.}\ \bibnamefont
  {Cantwell}}\ and\ \bibinfo {author} {\bibfnamefont {M.~E.~J.}\ \bibnamefont
  {Newman}},\ }\bibfield  {title} {\enquote {\bibinfo {title} {Message passing
  on networks with loops},}\ }\href {\doibase 10.1073/pnas.1914893116}
  {\bibfield  {journal} {\bibinfo  {journal} {Proceedings of the National
  Academy of Sciences}\ }\textbf {\bibinfo {volume} {116}},\ \bibinfo {pages}
  {23398--23403} (\bibinfo {year} {2019})},\ \Eprint
  {http://arxiv.org/abs/https://www.pnas.org/content/116/47/23398.full.pdf}
  {https://www.pnas.org/content/116/47/23398.full.pdf} \BibitemShut {NoStop}%
\bibitem [{\citenamefont {Hashimoto}(1989)}]{Hashimoto1989}%
  \BibitemOpen
  \bibfield  {author} {\bibinfo {author} {\bibfnamefont {K.}~\bibnamefont
  {Hashimoto}},\ }\bibfield  {title} {\enquote {\bibinfo {title} {Zeta
  functions of finite graphs and representations of p-adic g roups},}\ }\href
  {\doibase 10.1016/B978-0-12-330580-0.50015-X} {\bibfield  {journal} {\bibinfo
   {journal} {Adv. Stud. Pure Math.}\ }\textbf {\bibinfo {volume} {15}},\
  \bibinfo {pages} {211---280} (\bibinfo {year} {1989})}\BibitemShut {NoStop}%
\bibitem [{\citenamefont {Karrer}, \citenamefont {Newman},\ and\ \citenamefont
  {Zdeborov\'a}(2014)}]{Karrer2014}%
  \BibitemOpen
  \bibfield  {author} {\bibinfo {author} {\bibfnamefont {B.}~\bibnamefont
  {Karrer}}, \bibinfo {author} {\bibfnamefont {M.~E.~J.}\ \bibnamefont
  {Newman}}, \ and\ \bibinfo {author} {\bibfnamefont {L.}~\bibnamefont
  {Zdeborov\'a}},\ }\bibfield  {title} {\enquote {\bibinfo {title} {Percolation
  on sparse networks},}\ }\href {\doibase 10.1103/PhysRevLett.113.208702}
  {\bibfield  {journal} {\bibinfo  {journal} {Phys. Rev. Lett.}\ }\textbf
  {\bibinfo {volume} {113}},\ \bibinfo {pages} {208702} (\bibinfo {year}
  {2014})}\BibitemShut {NoStop}%
\bibitem [{\citenamefont {Saumell-Mendiola}, \citenamefont {\'{A}ngeles
  Serrano},\ and\ \citenamefont {Bogu\~{n}\'a}(2012)}]{Mendiola2012}%
  \BibitemOpen
  \bibfield  {author} {\bibinfo {author} {\bibfnamefont {A.}~\bibnamefont
  {Saumell-Mendiola}}, \bibinfo {author} {\bibfnamefont {M.}~\bibnamefont
  {\'{A}ngeles Serrano}}, \ and\ \bibinfo {author} {\bibfnamefont
  {M.}~\bibnamefont {Bogu\~{n}\'a}},\ }\bibfield  {title} {\enquote {\bibinfo
  {title} {Epidemic spreading on interconnected networks},}\ }\href {\doibase
  10.1103/PhysRevE.86.026106} {\bibfield  {journal} {\bibinfo  {journal} {Phys.
  Rev. E}\ }\textbf {\bibinfo {volume} {86}} (\bibinfo {year} {2012}),\
  10.1103/PhysRevE.86.026106}\BibitemShut {NoStop}%
\bibitem [{\citenamefont {Kivel\"a}\ \emph {et~al.}(2014)\citenamefont
  {Kivel\"a}, \citenamefont {Arenas}, \citenamefont {Barthelemy}, \citenamefont
  {Gleeson}, \citenamefont {Moreno},\ and\ \citenamefont
  {Porter}}]{Kivela2014}%
  \BibitemOpen
  \bibfield  {author} {\bibinfo {author} {\bibfnamefont {M.}~\bibnamefont
  {Kivel\"a}}, \bibinfo {author} {\bibfnamefont {A.}~\bibnamefont {Arenas}},
  \bibinfo {author} {\bibfnamefont {M.}~\bibnamefont {Barthelemy}}, \bibinfo
  {author} {\bibfnamefont {J.~P.}\ \bibnamefont {Gleeson}}, \bibinfo {author}
  {\bibfnamefont {Y.}~\bibnamefont {Moreno}}, \ and\ \bibinfo {author}
  {\bibfnamefont {M.~A.}\ \bibnamefont {Porter}},\ }\bibfield  {title}
  {\enquote {\bibinfo {title} {Multilayer networks},}\ }\href {\doibase
  10.1093/comnet/cnu016} {\bibfield  {journal} {\bibinfo  {journal} {J. Compl.
  Netw.}\ }\textbf {\bibinfo {volume} {2}} (\bibinfo {year} {2014}),\
  10.1093/comnet/cnu016}\BibitemShut {NoStop}%
\bibitem [{\citenamefont {Min}\ \emph {et~al.}(2016)\citenamefont {Min},
  \citenamefont {Gwak}, \citenamefont {Lee},\ and\ \citenamefont
  {Goh}}]{Min2016}%
  \BibitemOpen
  \bibfield  {author} {\bibinfo {author} {\bibfnamefont {B.}~\bibnamefont
  {Min}}, \bibinfo {author} {\bibfnamefont {S.-H.}\ \bibnamefont {Gwak}},
  \bibinfo {author} {\bibfnamefont {N.}~\bibnamefont {Lee}}, \ and\ \bibinfo
  {author} {\bibfnamefont {K.-I.}\ \bibnamefont {Goh}},\ }\bibfield  {title}
  {\enquote {\bibinfo {title} {Layer-switching cost and optimality in
  information spreading on multiplex networks},}\ }\href {\doibase
  10.1038/srep21392} {\bibfield  {journal} {\bibinfo  {journal} {Sci. Rep.}\
  }\textbf {\bibinfo {volume} {6}} (\bibinfo {year} {2016}),\
  10.1038/srep21392}\BibitemShut {NoStop}%
\end{thebibliography}%

%\bibliography{aipsamp}% Produces the bibliography via BibTeX.
\end{document}